\documentstyle[12pt]{article}

\begin{document}

\setcounter{page}{1}
\renewcommand{\thefootnote}{\fnsymbol{footnote}}
\pagestyle{plain} \vspace{1cm}

\begin{center}
\Large{\bf Black hole solutions in the warped DGP braneworld}\\
\small \vspace{1.5cm} {\bf Kourosh Nozari \footnote{knozari@umz.ac.ir}} \quad and
 \quad {\bf Amir Asaiyan \footnote{a.asaiyan@umz.ac.ir}}\\
\vspace{0.75 cm} {\it Department of Physics,\\
Faculty of Basic Sciences,\\
University of Mazandaran,\\
P. O. Box 47416-95447, Babolsar, IRAN}
\end{center}
\vspace{1.5cm}

\begin{abstract}
We study the static, analytical solution of black holes in the
warped DGP braneworld scenario. We show that the linearized field
equations and matching conditions lead to solutions that are not
compatible with Schwarzschild-(A)dS$_{(4)}$ solutions on the brane.
This incompatibility is similar to vDVZ discontinuity in massive
gravity theory. Following the standard procedure to remove this
discontinuity, which firstly was proposed by Vainshtein, we keep
some appropriate nonlinear terms in the field equations. This
strategy has its origin in the fact that the spatial extrinsic
curvature of the brane plays a crucial role in the nonlinear nature
of the solutions and also in recovering the well-measured
predictions of General Relativity (GR) at small scales. Using this
feature, we obtained an interesting black string solution in the
bulk when it is compatible with 4D GR solutions on the brane.\\
{\bf PACS}: 04.70.-s, 04.50.-h\\
{\bf Key Words}: Black Holes, Braneworlds, vDVZ Discontinuity.
\end{abstract}
\newpage

\section{\label{sec:1}Introduction}
A model universe with possible extra dimensions has attracted a lot
of attention in recent years. Questions such as: are we living
really in a four dimensional universe, or are there extra dimensions
we cannot see, have not been addressed precisely yet. But,
theoretically there are a lot of interesting outcomes by
incorporating some extra dimensions in theories such as gravitation.
In fact, extra dimensions are an integral part of fundamental
theories of physics such as superstring/M-theories that need more
than four spacetime dimensions. The idea to add an extra dimension
to General Relativity's four dimensional spacetime, firstly was
introduced by Kaluza and Klein to unify Electromagnetic and Gravity.
Now $p$-branes and $D$-branes are well-known in the context of
string theory where branes are solitonic solutions of
$10$-dimensional string theories. $p$-branes are extended objects of
higher dimension than strings ($1$-branes) and the $D$-branes are
kind of $p$-branes on which open strings can end \cite{mar}. In
braneworld scenarios we assume that our $(1+3)$-dimensional
spacetime is a domain wall embedded in a $5$-dimensional spacetime
called the bulk \cite{mar}. All matter fields and other gauge bosons
live on the brane but gravitons, which carry the gravitational
interaction, can travel into the extra dimension. There are at least
two basic reasons for investigating braneworld models: firstly for
solving hierarchy problem, that is, a large difference between
Planck and Electroweak scales, secondly for proposing a unified
theory of all fundamental interactions in the nature.

During the last two decades, different kinds of braneworld scenarios
have been proposed such as the Horava-Witten model \cite{hor},
Arkani Hamed-Dimopoulos-Dvali (ADD) model \cite{ark},
Randall-Sundrum (RS) models \cite{ran} and the
Dvali-Gabadadze-Porrati (DGP) model \cite{dgp}. Among these models,
one of which has attracted most regards in last decade, is the RS
one brane (RSII) model. In this model it is assumed that our
universe is a $3$-brane with positive tension, the extra dimension
is large and the bulk has a negative cosmological constant leading
to a warped geometry. Another model that has attractive results from
cosmological viewpoint, is the DGP model with one large extra
dimension but the bulk is Minkowski. In this model, there is a new
term in the total action that comes from quantum interaction between
matter confined on the brane and the bulk gravitons which induces
gravity on the brane. In this model, gravity leaks off the 4D brane
into the bulk and becomes 5-dimensional at large scales, $r\gg r_c$,
but on small scales, gravity is effectively bounded to the brane and
4D dynamics is regained. The transition from 4D to 5D behavior is
governed by the crossover scale $r_c$. An interesting property of
this model is that it contains a self-accelerating branch of the
solutions which can explain late time acceleration due to a
weakening of gravity at low energies \cite{deff}.

We note that a basic requirement for any alternative theory of
gravity, such as braneworld scenarios, is that they should reproduce
General Relativity predictions in the appropriate limit to be
phenomenologically viable. Two of the most important applications of
GR are cosmology and black holes. Here, our concentration is on the
issue of braneworld black holes, i.e. finding the bulk and the brane
metric when a spherically symmetric energy-momentum distribution is
localized on the brane. To obtain black hole solutions in a
braneworld scenario, we note that generally braneworld solutions can
be obtained by following two different approaches: In the first
approach, dynamics and geometry of the whole bulk spacetime are
primarily considered, then the dynamics on the brane is extracted by
using the Darmois-Israel matching conditions. The second approach is
to obtain effective 4-dimensional field equations on the brane
firstly and then try to extent these solutions to the bulk. This
approach has been applied to braneworlds with induced gravity in
Ref. \cite{mae}. The main difficulty in the later approach is that
it is not always possible to obtain a closed set of equations on the
brane. So, we are not able to predict the behavior of the fields on
the brane just with data on the brane.

In 1999, Chamblin, Hawking and Reall found the Schwarzschild black
string solution for RSII model as follows \cite{cha}
$${ds_{5}}^2=e^{\frac{-2|y|}{\ell}}{\tilde{g}}_{\mu\nu}dx^{\mu}dx^{\nu}+dy^{2}$$
$${\tilde{g}}_{\mu\nu}=e^{\frac{2|y|}{\ell}}g_{\mu\nu}=-\bigg(1-\frac{2GM}{r}\bigg)dt^{2}+
\frac{dr^{2}}{1-\frac{2GM}{r}}+r^{2}d\Omega^{2}\,.$$ In this
solution, there is a line singularity along $r=0$ for all $y$, but
the black string is unstable in response to large-scale
perturbations because the 5D curvature is unbounded at the Cauchy
horizon, as $y\rightarrow\infty$. Another solution for static
uncharged black hole was found by Dadhich  et al. from the induced
field equations on the brane \cite{dad}. In this work, a
Reissner-Nordstr\"{o}m type correction to the Schwarzschild solution
was found and 5D gravitational effects, which are impressed by the
bulk Weyl tensor, induce a tidal charge parameter $Q$. In this
approach the bulk metric has not been found. Braneworld black holes
have been studied extensively in recent years, some of these studies
can be found in Refs. \cite{cha}-\cite{mid}. An extension of the DGP
setup is the warped DGP scenario, which is a hybrid braneworld model
containing both RSII and the DGP models as its limits
\cite{mae,wDGP}. This model is constructed by considering the effect
of an induced gravity term in the RSII model. Black hole solutions
in a warped DGP braneworld have been studied recently in the case of
a conformally flat bulk for spherically symmetric vacuum on the
brane \cite{hey}. This study confirms the idea that an extra term in
the effective vacuum field equations on the brane can play the role
of a positive cosmological constant.

Within this streamline, in this paper we present an analytical
solution of static black hole in the warped DGP braneworld. Firstly,
we consider pure $(A)dS_{(5)}$ bulk field equations with a general
spherically symmetric 5D metric to achieve independent field
equations that should be held in the bulk. We take into account the
$3$-brane effect through appropriate matching conditions that act as
boundary conditions on 3-brane universe in order to constraint
general bulk solutions. In this way, we obtain a complicate system
of equations which should be solved to reach the warped DGP
braneworld, static, black hole solutions. Due to difficulties in
solving the field equations analytically, we consider the \emph{weak
field} limit of the scenario to find those solutions of the field
equations that describe the gravitational interactions in regions
far enough from the gravitational source. We demand these solutions
to support 4D General Relativity solutions for $r \ll r_c$ on the
brane. However, the full linearized theory that we have adopted,
does not lead to a correct 4D Schwarzschild-(A)dS$_{(4)}$ solution
on the brane. The reason for this incompatibility and its solution
are summarized in which follows: The linear analysis of the DGP
model shows that the tensor structure of the induced metric
perturbations takes the five dimensional form even at short
distances, i.e. in the limit $r_c\rightarrow \infty$, we have an
incompatibility with Einstein gravity \cite{dgp,deff,lue}. The
situation is analogous to the case of models with massive gravitons
in which we have a deviation from 4D GR (which is a massless
theory), even in the massless limit. This discontinuity in graviton
mass is known as \emph{van Dam-Veltman-Zakharov} (vDVZ)
discontinuity \cite{vdvz}. \emph{Vainshtein} claimed that the vDVZ
discontinuity is an artifact of the Pauli-Fierz Lagrangian, i.e. of
the linearization of the true, covariant, non-linear equations of
massive gravity, and the 4D GR is recovered by non-linear effect
\cite{vain}. There have been many discussions about this issue, see
for instance \cite{damour}. Another possibility that can disappear
the mentioned discontinuity is introducing a cosmological constant
\cite{mass}. In this way, in order to 4D GR be recovered, the
cosmological constant should be large enough as far as the graviton
mass is not completely negligibly small. The paradox in DGP gravity
seems to be that while it is clear that a perturbative, vDVZ-like
discontinuity occurs in the potential somewhere, no such
discontinuity appears in the cosmological solutions; at large Hubble
scales, the theory on the brane appears safely like general
relativity \cite{deff,lue}. Indeed, the cosmological solutions at
large Hubble scales are extremely nonlinear, and that perhaps, just
as Vainshtein suggested for massive gravity, nonlinear effects
become important in resolving the DGP version of the vDVZ
discontinuity. In the context of DGP model, there are various works
that indicate 4D GR is recovered at short distances
\cite{DGP2,gruz}. It was shown that the brane bending becomes
non-linear at scale shorter than $r_\ast\equiv(r_c^2r_g)^{1/3}$,
thus the linear analysis breaks down there. Also the leading order
correction to 4D GR has been obtained in this regime \cite{gruz}. In
Ref. \cite{tanaka}, an alternative formalism which can handle
general perturbations in weak gravity regime was proposed. They
generalized the DGP model by taking into account the bulk
cosmological constant and the brane tension balanced with it, so the
background metric is AdS$_{(5)}$ in the bulk and Minkowski on the
brane. By perturbing the background metric and following the prior
approach that had been proposed for RS scenario \cite{gariga}, they
confirmed the recovery of 4D GR at short distances and re-derived
the leading order correction to it. They have done this work by
taking into account the non-linear brane bending for weak gravity at
small scale $r<r_\ast$.

Here, in order to recover 4D Schwarzschild-(A)dS$_{(4)}$ solution on
the brane for $r<r_\ast$ or in the limit $r_c\rightarrow\infty$,
with regard to previous works, we take into account some nonlinear
terms in the field equations. In fact, these nonlinear terms become
so large in that regions that one cannot neglect them relative to
other nonlinear terms. We show that nonlinear effects become
important in resolving the "Warped DGP" version of the vDVZ
discontinuity too. By considering these nonlinear terms, we have
found interesting solutions for static Black holes on the warped DGP
brane.

\section{The field equations}

We consider a 3-brane $\Sigma$ embedded in a 5-dimensional bulk
${\cal{M}}$. The total action for the system is
\begin{equation}
S=\frac{1}{2\kappa_{5}^{2}}\int_{\cal
M}d^{5}\!X\sqrt{-g}\,\Big(^{(5)}\!R-2\Lambda_{5}\Big)+\int_{\cal
M}d^{5}\!X\sqrt{-g}\,\,{\cal L}_{bulk}+\,S_{brane}, \label{mat:1}
\end{equation}
where $S_{brane}$ is the 3-brane action defined as
\begin{equation}
S_{brane}=\frac{1}{2\kappa_{4}^{2}}\int_{\Sigma}d^{4}x\sqrt{-q}\,\Big(^{(4)}\!R-
2\Lambda_{4}\Big)+\int_{\Sigma}d^{4}x\sqrt{-q}\,{\cal
L}_{brane}+\int_{\Sigma}d^{4}x\sqrt{-q}\,\frac{K}{\kappa_{5}^{2}}\,.
\label{mat:2}
\end{equation}

$g_{AB}$ is the 5D bulk metric with corresponding Ricci tensor given
by $^{(5)}\!R_{AB}$. The brane has induced metric $q_{\mu\nu}$ with
corresponding Ricci tensor $^{(4)}\!R_{\mu\nu}$. The boundary
Gibbons-Hawking term is implied to yield the correct Einstein
equations in the bulk. $\Lambda_{5}$ and $\Lambda_{4}$ are bulk and
brane cosmological constants respectively.
$\frac{\Lambda_4}{\kappa_4^2}$ can be interpreted as the brane
tension of the standard Dirac-Nambu-Goto action and can include
quantum contributions to the four-dimensional cosmological constant
\cite{kof}. ${\cal{L}}_{bulk}$ and ${\cal {L}}_{brane}$ are the bulk
and brane matter Lagrangian respectively. We note also that the
Planck masses $M_5$ and $M_4$ are defined as $\kappa_5^2=8\pi
G_{(5)}=M_5^{-3}$ and $\kappa_4^2=8\pi G_{(4)}=M_4^{-2},$ where
$M_5$ and $M_4$ have dimensions of $(\rm length)^{-1}$.

We chose a coordinate $y$ for extra dimension so that our 3-brane in
the Gaussian Normal Coordinates is localized at $y=0$. By variation
of the action with respect to the bulk metric, 5D field equations
would be obtained as follows

\begin{equation} ^{(5)}\!G_{AB}=-\Lambda _{5}\, g_{AB}+\kappa
_5^{2}\,\,^{(bulk)}T_{AB}+\delta(y)\kappa _5^{2}\,\,^{(loc)}T_{AB},
\label{mat:3}
\end{equation}
where
\begin{equation}
^{(loc)}T_{AB}=g_A^{\mu}g_B^{\nu}\bigg(-\frac{1}{\kappa_4^2}\bigg)\sqrt{\frac{-q}{-g}}
\Big(^{(4)}\!G_{\mu\nu}+\Lambda_4\,
q_{\mu\nu}-\kappa_4^{2}\,\,^{(brane)}T_{\mu\nu}\Big), \label{mat:4}
\end{equation}
is the localized energy-momentum tensor on the brane and

\begin{equation}
^{(bulk)}T_{AB}=-2\frac{\delta{\cal L}_{bulk}}{\delta
g^{AB}}+g_{AB}{\cal L}_{bulk}\,. \label{mat:5}
\end{equation}

$^{(5)}G_{AB}$ and $^{(4)}G_{\mu\nu}$ denote the Einstein tensors
constructed from the bulk and the brane metrics respectively. The
tensor $q_{\mu\nu}$ is the induced metric on the brane $\Sigma$. The
field equations in the bulk take the following form
\begin{equation}
^{(5)}\!G_{AB}=^{(5)}\!\!\!R_{AB}-\frac{1}{2}\,^{(5)}\!R\,g_{AB}=-\Lambda_5g_{AB}+\kappa_5^{2}\,\,^{(bulk)}T_{AB},
\label{mat:6}
\end{equation}
or
\begin{equation}
^{(5)}\!R_A^{\,B}-\frac{1}{2}\,^{(5)}\!R\,\delta_A^{\,B}=-\Lambda_5\delta_A^{\,B}+\kappa_5^{2}\,\,^{(bulk)}T_A^{\,B}.
\label{mat:7}
\end{equation}

Moreover, the following modified (due to the presence of induced
gravity on the brane) Israel-Darmois junction conditions would be
obtained from the distributional character of the brane content
\begin{equation}
[K_{\mu\nu}]-q_{\mu\nu}[K]=-\kappa_5^{2}\,\,^{(loc)}T_{\mu\nu}=
\frac{\kappa_5^2}{\kappa_4^2}\Big(^{(4)}\!G_{\mu\nu}+\Lambda_4q_{\mu\nu}\Big)-\kappa_5^{2}\,\,^{(brane)}T_{\mu\nu},
\label{mat:8}
\end{equation}
where $K_{\mu\nu}=\frac{1}{2}\partial_y(g_{\mu\nu})$ is the
extrinsic curvature of the brane and brackets denote jump across the
brane $(y=0)$.

Now we consider a general, 5-dimensional, static metric with
spherical symmetry on the brane as follows
\begin{equation}
{ds_5}^2=-e^{\nu(r,y)}dt^2+e^{\lambda(r,y)}dr^2+r^2e^{\mu(r,y)}(d\theta^2+\sin^{2}\theta
d\phi^2)+dy^2\,. \label{mat:9}
\end{equation}

Also we assume the matter content of the bulk is just a cosmological
constant $\Lambda_5$ (i.e., $^{(bulk)} T_{AB}=0$) and the matter
content of the 3-brane universe is considered to be a cosmological
constant $\Lambda_4$ plus a localized spherically symmetric untilted
perfect fluid (e.g. a star) with
\begin{equation}
^{(brane)}T_{\mu\nu}=(\rho+p)u_\mu u_\nu+pq_{\mu\nu}\,,
\label{mat:10}
\end{equation}
where $u^\mu$ stands for the 4-velocity of the fluid and $\rho=p=0$
for $r>R$. Nevertheless, since we want to obtain static black hole
solutions outside the star ( that is, for $r>R$), in these regions
the brane has just a cosmological constant $\Lambda_4$.

Hereafter, we consider a ${\bf Z}_2$-symmetry on reflection across
the brane, thus at $r>R$ the Israel matching conditions become

\begin{equation}
\overline{K}_\mu^{\,\nu}-\overline{K}\delta_\mu^{\,\nu}=r_c\Big(^{(4)}\!G_\mu^\nu+\Lambda_4\delta_\mu^\nu\Big),
\label{mat:11}
\end{equation}
where $r_c=\frac{\kappa_5^2}{2\kappa_4^2}=\frac{M_4^2}{2M_5^3}$ is
the DGP crossover distance, and by definition
$\overline{K}_\mu^{\,\nu}=K_\mu^{\,\nu}(y=0^+)=-K_\mu^{\,\nu}(y=0^-)$.
Now by putting the metric (\ref{mat:9}) into the bulk field
equations (\ref{mat:7}) we obtain the $(tt)$,\, $(rr)$,\,
$(\theta\theta)$\, and\, $(ry)$\, components of the bulk field
equations respectively as follows

$$4r\mu_{rr}+3r\mu_{r}^2+12\mu_r-4\lambda_r-2r\lambda_r\mu_r+\frac{4}{r}\Big(1-e^{\lambda-\mu}\Big)+$$
\begin{equation}
2r\lambda_{yy}e^{\lambda}+4r\mu_{yy}e^{\lambda}+r\lambda_{y}^2e^{\lambda}+3r\mu_{y}^2e^{\lambda}+
2r\lambda_y\mu_ye^{\lambda}=4r\Lambda_5e^{\lambda}\,, \label{mat:12}
\end{equation}

$$r\mu_{r}^2+2r\nu_r\mu_r+4\nu_r+4\mu_r+\frac{4}{r}\Big(1-e^{\lambda-\mu}\Big)+2r\nu_{yy}e^{\lambda}+$$
\begin{equation}
4r\mu_{yy}e^{\lambda}+r\nu_{y}^2e^{\lambda}+3r\mu_{y}^2e^{\lambda}+2r\nu_y\mu_ye^{\lambda}=4r\Lambda_5e^{\lambda}\,,
\label{mat:13}
\end{equation}

$$2r\nu_{rr}+2r\mu_{rr}+r\nu_{r}^2+r\mu_{r}^2-r\nu_r\lambda_r+r\nu_r\mu_r-r\lambda_r\mu_r+2\nu_r-2\lambda_r+4\mu_r+$$
\begin{equation}
2r\big[\nu_{yy}+\lambda_{yy}+\mu_{yy}\big]e^{\lambda}+r\big[\lambda_y\nu_y+\lambda_y\mu_y+\nu_y\mu_y\big]e^{\lambda}+
r\big[\lambda_y^2+\nu_{y}^2+\mu_{y}^2\big]e^{\lambda}=4r\Lambda_5e^{\lambda}\,,
\label{mat:14}
\end{equation}

$$4r\mu_{rr}+2r\nu_{rr}+3r\mu_{r}^2+r\nu_{r}^2+2r\nu_r\mu_r-2r\lambda_r\mu_r-r\nu_r\lambda_r+4\nu_r-4\lambda_r+12\mu_r+$$
\begin{equation}
\frac{4}{r}\Big(1-e^{\lambda-\mu}\Big)+r\mu_{y}^2e^{\lambda}+r\big[\lambda_y\nu_y+2\lambda_y\mu_y+4\nu_y\mu_y\big]e^{\lambda}=4r\Lambda_5e^{\lambda}\,,
\label{mat:15}
\end{equation}

\begin{equation}
2\nu_{ry}+4\mu_{ry}+\nu_r\nu_y-\nu_r\lambda_y+2\mu_r\mu_y-2\lambda_y\mu_r+\frac{4}{r}\Big(\mu_y-\lambda_y\Big)=0\,.
\label{mat:16}
\end{equation}

By using the Israel matching conditions (\ref{mat:11}), we obtain
the following conditions on the brane

\begin{equation}
-\frac{1}{2}\Big(2\mu_y+\lambda_y\Big)|_{y=0^+}=r_c\Big\{\Lambda_4-\frac{e^{-\lambda}}{4r^2}
\Big(4-4e^{\lambda-\mu}+4r^2\mu_{rr}+3r^2\mu_{r}^2-4r\lambda_r+12r\mu_r-2r^2\lambda_r\mu_r\Big)\Big\}\,,
\label{mat:17}
\end{equation}

\begin{equation}
-\frac{1}{2}\Big(2\mu_y+\nu_y\Big)|_{y=0^+}=r_c\Big\{\Lambda_4-\frac{e^{-\lambda}}{4r^2}
\Big(4-4e^{\lambda-\mu}+r^2\mu_{r}^2+4r\mu_r+4r\nu_r+2r^2\nu_r\mu_r\Big)\Big\}\,,
\label{mat:18}
\end{equation}

$$-\frac{1}{2}\Big(\nu_y+\lambda_y+\mu_y\Big)|_{y=0^+}=r_c\Big\{\Lambda_4-\frac{e^{-\lambda}}{4r}\Big(2r\nu_{rr}+
2r\mu_{rr}+r\mu_{r}^2+r\nu_{r}^2$$
\begin{equation}
\hspace{7cm}-r\nu_r\lambda_r+2\nu_r-2\lambda_r+4\mu_r+r\nu_r\mu_r-r\lambda_r\mu_r
\Big)\Big\}\,. \label{mat:19}
\end{equation}

The subscripts $y$ and $r$ in these relations represent partial
differentiation with respect to $y$ and $r$ respectively. Note that
these equations are held on the brane outside our spherical object.

\section{The weak field solutions}

The nonlinear, second order partial differential equations
(\ref{mat:12})\,-\,(\ref{mat:16}) should be solved and the
corresponding solutions should satisfy junction conditions
(\ref{mat:17}), (\ref{mat:18}) and (\ref{mat:19}) across the ${\bf
Z}_2$-symmetric brane. This is not an easy task at all! To find some
analytical solutions, we consider just the \emph{weak field} regime
(i.e., far enough from the source localized on the brane). In this
respect, we adopt the assumption that $|\nu|$,\, $|\lambda|$ and
$|\mu|$ are small quantities compared to unity; that is, $|\nu|,\,
|\lambda|,\, |\mu|\ll 1$. By adopting this assumption, we linearize
our field equations and also brane boundary conditions. By keeping
only the leading order terms, the bulk equations
(\ref{mat:12})\,-\,(\ref{mat:16}) reduce to the following equations

\begin{equation}
2r\mu_{rr}+4\mu_r-2\lambda_r-\nu_r-\frac{c}{r^2}+r(\lambda_{yy}+2\mu_{yy})=\frac{4}{3}r\Lambda_5\,,
\label{mat:20}
\end{equation}

\begin{equation}
\nu_r-\frac{c}{r^2}+r(\nu_{yy}+2\mu_{yy})=\frac{4}{3}r\Lambda_5\,,
\label{mat:21}
\end{equation}

\begin{equation}
r(\nu_{rr}+\mu_{rr})+\nu_r+2\mu_r-\lambda_r+r(\nu_{yy}+\lambda_{yy}+\mu_{yy})=2r\Lambda_5\,,
\label{mat:22}
\end{equation}

\begin{equation}
\frac{2}{r}(\lambda-\mu)=\nu_r+2\mu_r+\frac{c}{r^2}-\frac{2}{3}r\Lambda_5\,,
\label{mat:23}
\end{equation}
where $c$ is an integration constant to be fixed later. In the same
procedure applied to equations (\ref{mat:17}), (\ref{mat:18}) and
(\ref{mat:19}), we achieve the following linearized junction
conditions

\begin{equation}
-\frac{1}{2}\big(\lambda_y+2\mu_y\big)|_{y=0^+}=r_c\Big\{\Lambda_4-\frac{1}{r^2}\Big(\mu-\lambda+3r\mu_r+r^2\mu_{rr}-r\lambda_r\Big)\Big\},
\label{mat:24}
\end{equation}

\begin{equation}
-\frac{1}{2}\big(\nu_y+2\mu_y\big)|_{y=0^+}=r_c\Big\{\Lambda_4-\frac{1}{r^2}\Big(\mu-\lambda+r\mu_r+r\nu_r\Big)\Big\},
\label{mat:25}
\end{equation}

\begin{equation}
-\frac{1}{2}\big(\nu_y+\lambda_y+\mu_y\big)|_{y=0^+}=r_c\Big\{\Lambda_4-\frac{1}{2r}\Big(r\nu_{rr}+r\mu_{rr}+2\mu_r+\nu_r-\lambda_r\Big)\Big\}.
\label{mat:26}
\end{equation}

By solving the bulk field equations
(\ref{mat:20})\,-\,(\ref{mat:23}), we find $\mu$ and $\lambda$ in
terms of $\nu$ as follows

\begin{equation}
\mu(r,y)=-\frac{1}{2r}\int\int\nu_rdydy-\frac{1}{2}\nu+\frac{c}{4r^3}y^2+\frac{1}{3}\Lambda_5y^2+\bigg(F_1(r)+H(r)\bigg)|y|+F_2(r)+G(r)\,,
\label{mat:27}
\end{equation}

$$\lambda(r,y)=-2\int\int\nu_{rr}dydy-\frac{3}{r}\int\int\nu_rdydy-2\nu+\frac{c}{2r}+\frac{2}{3}\Lambda_5r^2+r|y|F'_1(r)+$$
\begin{equation}
|y|F_1(r)+rF'_2(r)+F_2(r)+\frac{4}{3}\Lambda_5y^2-\frac{c}{2r^3}y^2+A(r)|y|+B(r)\,,
\label{mat:28}
\end{equation}
where $A(r)$, $B(r)$, $F_1(r)$, $F_2(r)$, $G(r)$ and $H(r)$ are
arbitrary functions of $r$,\, and a prime denotes derivative with
respect to $r$. Also $\nu(r,y)$ should satisfy the following
differential equation

\begin{equation}
\nu_{rryy}=-\nu_{yyyy}-\frac{2}{r^3}\nu_r-\frac{1}{r}\nu_{ryy}+\frac{1}{r}\nu_{rrr}+\frac{2}{r^2}\nu_{rr}\,.
\label{mat:29}
\end{equation}

Therefore, for a given $\nu(r,y)$ that satisfies this equation,
$\mu(r,y)$ and $\lambda(r,y)$ can be obtained via (\ref{mat:27}) and
(\ref{mat:28}) respectively. Equation (\ref{mat:29}) has the
following solution for $\nu(r,y)$

\begin{equation}
\nu(r,y)=\frac{a}{r}+br^2+ey^2+d|y|+\frac{f}{r}|y|+qr^2|y|\,.
\label{mat:30}
\end{equation}

Now by using equations (\ref{mat:27}) and (\ref{mat:28}), we
calculate $\lambda(r,y)$ and $\mu(r,y)$ to find

$$\lambda(r,y)=\Big(-\frac{f}{6r^3}-\frac{5}{3}q\Big)|y|^3-\Big(\frac{(a+c)}{2r^3}+\frac{b}{2}+2e-\frac{4}{3}\Lambda_5\Big)y^2-$$
\begin{equation}
\Big(2qr^2+\frac{2}{r}f+2d-rF'_1-F_1-P(r)\Big)|y|+\frac{c-4a}{2r}-2br^2+\frac{2}{3}\Lambda_5r^2+rF'_2+F_2+Q(r)\,,
\label{mat:31}
\end{equation}

$$\mu(r,y)=\Big(\frac{f}{12r^3}-\frac{q}{6}\Big)|y|^3+\Big(\frac{(a+c)}{4r^3}+\frac{1}{3}\Lambda_5+\frac{b}{4}-\frac{e}{2}\Big)y^2-$$
\begin{equation}
\Big(\frac{f}{2r}+\frac{q}{2}r^2+\frac{d}{2}-F_1-M(r)\Big)|y|-\frac{a}{2r}-\frac{b}{2}r^2+F_2+N(r)\,,
\label{mat:32}
\end{equation}
where $a$, $b$, $d$, $e$, $f$ and $q$ are constants, whereas
$P(r)$,\, $Q(r)$,\, $M(r)$ and $N(r)$ are arbitrary functions of
$r$. Equations (\ref{mat:20})\,-\,(\ref{mat:26}) can help us to
determine the arbitrary parameters and functions appeared in
$\nu(r,y)$, $\lambda(r,y)$ and $\mu(r,y)$. By this way, we found
that $b$, $d$, $f$ and $q$ should be zero and also
$\Lambda_5=-\Lambda_4$ and $e=\frac{2}{3}\Lambda_5$ in order for
equations (\ref{mat:20})\,-\,(\ref{mat:26}) to be satisfied. In this
situation the solutions simplify to the following equations

\begin{equation}
\nu(r,y)=\frac{a}{r}+\frac{2}{3}\Lambda_5y^2, \label{mat:33}
\end{equation}

\begin{equation}
\lambda(r,y)=-\Big(\frac{a+c}{2r^3}\Big)y^2+r_c\Big(\frac{4}{3}\Lambda_5+\frac{a+c}{r^3}\Big)|y|+\frac{c-a}{2r}-\frac{1}{3}\Lambda_5r^2+rW'+W(r),
\label{mat:34}
\end{equation}

\begin{equation}
\mu(r,y)=\Big(\frac{a+c}{4r^3}\Big)y^2+r_c\Big(\frac{4}{3}\Lambda_5-\frac{a+c}{2r^3}\Big)|y|+W(r)-\frac{a}{2r}\,,
\label{mat:35}
\end{equation}
where $W(r)=N(r)+F_2(r)$. The corresponding quantities on the brane
($y=0$) take the following forms

\begin{equation}
\nu(r,y=0)=\frac{a}{r}, \label{mat:36}
\end{equation}

\begin{equation}
\lambda(r,y=0)=\frac{c-a}{2r}+\frac{1}{3}\Lambda_4r^2+rW'+W(r),
\label{mat:37}
\end{equation}

\begin{equation}
\mu(r,y=0)=W(r)-\frac{a}{2r}. \label{mat:38}
\end{equation}

Since we expect $\mu(r,y=0)=0$, we set $W(r)=\frac{a}{2r}$. So we
find

\begin{equation}
\lambda(r,y=0)=\frac{c-a}{2r}+\frac{1}{3}\Lambda_4r^2\,.
\label{mat:39}
\end{equation}

$c$\, is an integration constant which we set to be zero in order to
find more familiar results as follows

\begin{equation}
\nu(r,y)=\frac{a}{r}+\frac{2}{3}\Lambda_5y^2, \label{mat:40}
\end{equation}

\begin{equation}
\lambda(r,y)=-\frac{a}{2r}+\frac{1}{3}\Lambda_4r^2-\frac{a}{2r^3}y^2+r_c\Big(\frac{4}{3}\Lambda_5+\frac{a}{r^3}\Big)|y|\,,
\label{mat:41}
\end{equation}

\begin{equation}
\mu(r,y)=\frac{a}{4r^3}y^2+r_c\Big(\frac{4}{3}\Lambda_5-\frac{a}{2r^3}\Big)|y|\,.
\label{mat:42}
\end{equation}

Note that in these relations we have used $\Lambda_5=-\Lambda_4$
interchangeably. To find Schwarzschild solution in appropriate
limit, we set $a\!=\!-2m$\,, where $m$ is the Schwarzschild
geometric mass. It is clear that in our linearized warped DGP setup,
$\lambda(r,y=0)\neq -\nu(r,y=0)$ on the 3-brane. We expect that in
the limit of $r_c\!\!\rightarrow\!\!\infty$\, the solutions reduce
to the 4D general relativistic solutions on the brane, that is, to a
Schwarzschild-(A)dS$_{(4)}$ metric. However, this has not occurred
here. We note that the reason behind having a
non-Schwarzschild-(A)dS$_{(4)}$ solution can be explained through
the fact that the curvatures of the bulk and brane spacetimes should
not exactly cancel each other. In the next section we focus on this
issue.

\section{Recovering the Schwarzschild solution on the brane}

To recover the Schwarzschild-(A)dS$_{(4)}$ solution on the brane, we
note that all metric components should be much smaller than unity as
a result of the weak field prescription. But this is violated by the
linearized solutions we obtained in the last section. Nevertheless,
if we look at the 5D bulk solutions Eqs.~(\ref{mat:40}),
(\ref{mat:41}) and (\ref{mat:42}), we will see that in 5D, $\lambda$
and $\mu$ are large for large $r_c$,\,( see for instance
\cite{gruz}, \cite{DGP2} and \cite{lue}). Although we are interested
only in 4D solutions, the linearized theory is not applicable at
large $r_c$\,. These features show that the completely linearized
theory does not result in correct 4D Schwarzschild-(A)dS$_{(4)}$
solutions (which are given by $\mu(r)=0$\,, $\lambda(r)=-\nu(r)$ and
$\nu(r)=-\frac{2m}{r}-\frac{1}{3}\Lambda_4r^2$). In fact, in the
limit $r_c\!\!\rightarrow\!\!\infty$,\, but yet in the weak field
regime, we cannot neglect all nonlinear terms in some subspaces. For
instance, if $r_c^2|a|\gg r^3$\,, quantities such as
$\lambda_y$\,and\, $\mu_y$ are large enough that we have to save
their squares in the field equations. Therefore, for $|a|\ll r\ll
r_*\equiv(r_c^2|a|)^{\frac{1}{3}}$ the nonlinear field equations
that should be solved are as follows\footnote{We note that a static
source placed on a brane creates a nonzero scalar curvature around
it. For a source of the size $<r_\ast$\,, this curvature extends to
a distance $\sim r_\ast$. More intuitively, a static source distorts
a brane medium around it creating a potential well, and the
distortion extends to a distance $r\sim r_\ast$. This curvature
suppresses nonlinear interactions that otherwise would become strong
at the scale below $r_\ast$\,, see for instance \cite{gab}.}

\begin{equation}
\frac{2}{r}\Big(\mu-\lambda\Big)=-\nu_{r}-2\mu_{r}+U(r)\,,
\label{mat:43}
\end{equation}

\begin{equation}
4r\mu_{rr}+8\mu_r-4\lambda_r-2\nu_r+2U(r)+2r(\lambda_{yy}+2\mu_{yy})+r\lambda_{y}^2+3r\mu_{y}^2+
2r\lambda_y\mu_y=4r\Lambda_5\,, \label{mat:44}
\end{equation}

\begin{equation}
2\nu_r+2U(r)+2r\nu_{yy}+4r\mu_{yy}+3r\mu_{y}^2=4r\Lambda_5\,,
\label{mat:45}
\end{equation}

\begin{equation}
2r\nu_{rr}+2r\mu_{rr}+2\nu_r-2\lambda_r+4\mu_r+2r\big(\nu_{yy}+\lambda_{yy}+\mu_{yy}\big)+r\lambda_y\mu_y+
r\big(\lambda_y^2+\mu_{y}^2\big)=4r\Lambda_5\,, \label{mat:46}
\end{equation}

\begin{equation}
4r\mu_{rr}+2r\nu_{rr}+2\nu_r-4\lambda_r+8\mu_r+2U(r)+2r\lambda_y\mu_y+r\mu_{y}^2=4r\Lambda_5\,,
\label{mat:47}
\end{equation}
where $U(r)$ is an arbitrary function of $r$\,. Junction conditions
in this situation are the same as the linearized case given by Eqs.
(\ref{mat:24}), (\ref{mat:25}) and (\ref{mat:26}). By solving this
set of equations, we find the following class of solutions

\begin{equation}
\nu(r,y)=\frac{a}{r}+\frac{1}{9}\Lambda_{5}r^{2}+\frac{4}{9}r_{c}\Lambda_{5}|y|+\frac{4}{27}r_c^2\Lambda_5^2y^{2}\,,
\label{mat:48}
\end{equation}
\begin{equation}
\lambda(r,y)=-\frac{a}{r}-\frac{1}{9}\Lambda_{5}r^{2}+\frac{4}{9}r_{c}\Lambda_{5}|y|\,,
\label{mat:49}
\end{equation}
\begin{equation}
\mu(r,y)=\frac{4}{9}r_{c}\Lambda_{5}|y|\,, \label{mat:50}
\end{equation}
where now $\Lambda_4=-\frac{1}{3}\Lambda_5$\, with
\,$\Lambda_4=-\frac{3}{4}\frac{1}{r_c^2}$\,. Note that in this
framework we obtained a geometric interpretation of the brane
cosmological constant in terms of the DGP crossover scale. Moreover,
as we have mentioned at the end of the previous section, the
curvatures of the bulk and brane spacetimes do not cancel each other
exactly which is the reason behind having a
non-Schwarzschild-(A)dS$_{(4)}$ solution in the previous section. On
the brane with $y=0$, we find

\begin{equation}
\nu(r,y=0)=\frac{a}{r}-\frac{1}{3}\Lambda_{4}r^{2}\,, \label{mat:51}
\end{equation}
\begin{equation}
\lambda(r,y=0)=-\frac{a}{r}+\frac{1}{3}\Lambda_{4}r^{2}\,,
\label{mat:52}
\end{equation}
\begin{equation}
\mu(r,y=0)=0\,, \label{mat:53}
\end{equation}
which is compatible with 4D Schwarzschild-AdS$_{(4)}$ solution at
the leading order if  we set $a\!\equiv\!-2m$
\begin{equation}
ds^{2}=-\bigg(1-\frac{2m}{r}-\frac{1}{3}\Lambda_{4}r^{2}\bigg)dt^{2}+
\frac{dr^{2}}{\bigg(1-\frac{2m}{r}-\frac{1}{3}\Lambda_{4}r^{2}\bigg)}+r^{2}d\Omega^{2}\,.
\label{mat:54}
\end{equation}

Therefore, the Einstein gravity is recovered and the vDVZ problem is
resolved by this strategy. Indeed, the spatial extrinsic curvatures
of the brane, i.e. $\lambda_y|_{y=0}$\, and \,$\mu_y|_{y=0}$\,, play
a crucial role in the nonlinear nature of the solution and
recovering the predictions of general relativity. In this respect,
the nonlinear behavior arises from purely spatial geometric factors
(see also Ref. \cite{gruz}). We note also that there is a massless
scalar mode, which arises from decomposition of five degrees of
freedom of the bulk graviton in the massless limit (i.e.
$r_c\!\!\rightarrow\!\!\infty$). This extra scalar mode persists as
an extra degree of freedom in all regimes of the theory. The
extrinsic curvature suppresses this extra scalar field inside the
region $r\ll r_*$, whereas outside this region, the brane bending
mode is free to propagate (see Ref. \cite{lue} for more details).

We note that Eqs. (\ref{mat:48}), (\ref{mat:49}) and (\ref{mat:50})
can be rewritten by using $\Lambda_4=-\frac{3}{4}\frac{1}{r_c^2}$ as
follows

\begin{equation}
\nu(r,y)=-\frac{2m}{r}+\frac{1}{4}\frac{r^{2}}{r_c^2}+\frac{1}{r_c}|y|+
\frac{3}{4}\frac{y^2}{r_c^2}\,, \label{mat:55}
\end{equation}
\begin{equation}
\lambda(r,y)=\frac{2m}{r}-\frac{1}{4}\frac{r^{2}}{r_c^2}+\frac{1}{r_c}|y|\,,
\label{mat:56}
\end{equation}
\begin{equation}
\mu(r,y)=\frac{1}{r_c}|y|\,. \label{mat:57}
\end{equation}

In the limit $r_{c}\!\rightarrow\! \infty$ we find the Schwarzschild
solution on the brane. This means that the effect of the brane
cosmological constant is negligible in this limit. Thus, it is
essentially impossible to observe a Schwarzschild-(A)dS$_{(4)}$
interaction in our observations, and the Newtonian gravity is
dominant at least in our solar system scale. In is important to note
that the above solution is actually a black string solution since we
find a singularity at $r=0$ for any value of $y$. This can be shown
through singularity of the 5-dimensional Kretschmann scalar. The
expression of the 5-dimensional Kretschmann scalar is to long to be
presented here, but our calculation by using the Maple package, has
shown that the 5-dimensional Kretschmann scalar is singular at $r=0$
for all values of $y$. In fact, in the $r\rightarrow0$ limit, it
reduces to the following simpler form
\begin{equation}
\lim _{\quad\quad
r\rightarrow0}K^2=\frac{32r_c^3y+16r_c^2y^2+16r_c^4}{(r_c+y)^4}\frac{1}{r^4}
\end{equation}
which obviously diverges at $r=0$. This shows that singularity at
$r=0$ is an intrinsic singularity, and this is true for all values
of $y$, which confirm that the solutions are actually black string.
Moreover, the 4-dimensional Kretschmann scalar of the metric
(\ref{mat:54}) expressed in terms of $r_{c}$ is given by
\begin{equation}
K^2\equiv\,^{(4)}\!R_{\alpha\beta\gamma\delta}\,^{(4)}\!R^{\alpha\beta\gamma\delta}=
\frac{3}{2}\frac{\big(\frac{r^6}{r_c^4}+32m^2\big)}{r^6}\,,
\label{mat:58}
\end{equation}
which shows that singularity at $r=0$ on the brane is an intrinsic
singularity. In the limit of $r_c\!\!\rightarrow\!\!\infty$, we have
just one horizon which is the Schwarzschild horizon at $r=2m$. But
for $m=0$ there is no de Sitter horizon on the brane. The stability
of black string solutions should be considered in the light of the
results of Gregory-Laflamme pioneering work \cite{gr}.

Another issue which requires especial attention here is the fact
that a de Sitter bulk invalidates most of the attractive points of
the Randall-Sundrum model such as the localization of the zero mode
of the graviton near the brane. The question then arises: how is the
situation in the warped DGP setup? In DGP-like models that have a 4D
Ricci scalar in the bulk action, gravity becomes 5-dimensional at
$r\gg r_c$ where 5D Einstein-Hilbert action is dominant. For $r\ll
r_c$ gravity is 4-dimensional but not the 4D GR one. This property
of these models is definitely different from RSII model, in which at
low energies, $H\ell\ll 1$, 4D gravity is recovered to a good
approximation. In RSII model what prevents gravity from leaking into
the extra dimension at low energies is the negative bulk
cosmological constant. Moreover, in RSII model the 4D gravitational
constant strongly relies on the presence of the brane tension, and
the brane tension should be positive to effective 4D gravitational
constant have correct sign. This positive tension and the bulk
negative cosmological constant are balanced with each other to have
a zero 4D cosmological constant on the brane. In our warped DGP
setup, we found $\Lambda_4<0$ and $\Lambda_5>0$, which is completely
different from the RSII case. Here we don't need a negative
$\Lambda_5$ and positive $\Lambda_4$ because induced gravity term
gives us the desired results. Solutions that we have found in Sec 4,
Eqs. (55)-(57), are valid only at $r\ll r_\ast$ and for $r\gg r_c$
appropriate solutions should be obtained. Nevertheless, since a
warped DGP scenario is a hybrid braneworld model which contains both
the pure DGP and RSII models in appropriate limits \cite{mae}, it is
expected that some of the mentioned shortcomings are still present
in this case too. This needs further justification that we are going
to study separately.

As the final remark, we note that Birkhoff's theorem is absent in
theories of modified gravity such as the DGP braneworld
scenario\cite{dai2}. By calculation of the gravitational force on a
test particle due to a spherical mass shell in the DGP setup, one
can show that unlike in GR, the force depends on the mass
distribution. In particular, the gravitational force within a
spherical mass shell depends on the geometric structure of the bulk.
In our setup, having Schwarzschild-(A)dS$_{(4)}$ solution means that
Birkhoff's theorem holds good and this is due to geometric structure
of the bulk manifold in this case which differs from pure DGP case.

\section{Summary}

In this paper we obtained a class of static black hole solutions in
the warped DGP braneworld. Firstly we solved the general bulk field
equations in the weak field limit (by linearizing the field
equations and matching conditions). The solutions on the brane then
were found by using the Israel matching conditions. However, these
brane solutions were not compatible with Schwarzschild-(A)dS$_{(4)}$
solutions on the brane. We adopted a strategy to solve this problem
based on keeping appropriate nonlinear terms in the field equations.
This strategy has its origin in the fact that the spatial extrinsic
curvature of the brane plays a crucial role in the nonlinear nature
of the solutions, and also in recovering the predictions of General
Relativity. In fact, the nonlinear behavior arises from purely
spatial geometric factors. Using this feature, we obtained some
interesting black sting solutions compatible with well-known black
hole solutions on the brane.\\


\begin{thebibliography}{99}
\bibitem{mar}
C. Csaki, \textit {TASI Lectures on Extra Dimensions and Branes }, [arXiv:hep-ph/0404096].\\
R. Maartens and K. Koyama,\textit{Brane-World Gravity}, Living Rev.
Relativity {\bf13} (2010) 5, [arXiv:1004.3962].

\bibitem{hor}
P. Horava and E. Witten, \textit {Hetrotic and Type I String
Dynamics from Eleven Dimensions},  Nucl. Phys. B {\bf460} (1996)
506, [arXiv:hep-th/9510209].

\bibitem{ark}
N. Arkani-Hamed, S. Dimopoulos and G. Dvali, \textit {The Hierarchy
Problem and New Dimensions at a Millimeter}, Phys. Lett. B {\bf429}
(1998) 263, [arXiv:hep-ph/9803315].

\bibitem{ran}
L. Randall and R. Sundrum, \textit {A Large Mass Hierarchy from a
Small Extra Dimension}, Phys. Rev. Lett. {\bf83} (1999) 3370, [arXiv:hep-ph/9905221].\\
L. Randall and R. Sundrum, \textit {An Alternative to
Compactification}, Phys. Rev. Lett. {\bf83} (1999) 4690,
[arXiv:hep-th/9906064].

\bibitem{dgp}
G. Dvali, G. Gabadadze and M. Porrati, \textit {4D Gravity on a
Brane in 5D Minkowski Space}, Phys. Lett. B {\bf485} (2000) 208,
[arXiv:hep-th/0005016].

\bibitem{mae}
K. -i. Maeda, S. Mizuno and T. Torii, \textit {Effective
gravitational equations on a brane world with induced gravity},
Phys. Rev. D {\bf68} (2003) 024033, [arXiv:gr-qc/0303039].

\bibitem{wDGP}
H. Collins and B. Holdom, \textit {Brane cosmologies without orbifolds}, Phys. Rev. D \textbf{62}, (2000) 105009.\\
Y. V. Shtanov, \textit {On Brane-World Cosmology}, [arXiv:hep-th/0005193].\\
V. Sahni and Y. Shtanov, \textit {Braneworld models of dark energy},
[arXiv:astro-ph/0202346].

\bibitem{deff}
C. Deffayet, \textit {Cosmology on a brane in Minkowski bulk}, Phys.
Lett. B {\bf502}, (2001) 199, [arXiv:hep-th/0010186].\\
G. R. Dvali and G. Gabadadze, \textit {Gravity on a brane in infinite-volume extra space},  Phys. Rev. D \textbf{63}, (2001) 065007, [arXiv:hep-th/0008054].\\
C. Deffayet, G. R. Dvali and G. Gabadadze, \textit {Accelerated
universe from gravity leaking to extra dimensions} , Phys. Rev. D
\textbf{65}, (2002) 044023, [arXiv:astro-ph/0105068].

\bibitem{lue}
A. Lue, \textit {The phenomenology of Dvali-Gabadadze-Porrati
cosmologies}, Phys. Rept. {\bf423} (2006) 1-48,
[arXiv:astro-ph/0510068].

\bibitem{cha}
A. Chamblin, S. W. Hawking and H. S. Reall, \textit {Brane-World
Black Holes}, Phys. Rev. D {\bf61} (2000) 065007,
[arXiv:hep-th/9909205].

\bibitem{dad}
N. Dadhich, R. Maartens, P. Papadopoulos and V. Rezania, \textit
{Black holes on the brane}, Phys. Lett. B {\bf487} (2000) 1-6,
[arXiv:hep-th/0003061].

\bibitem{kof}
G. Kofinas, E. Papantonopoulos and V. Zamarias, \textit {Black Hole
Solutions in Braneworlds with Induced Gravity}, Phys. Rev. D
{\bf66} (2002) 104028, [arXiv:hep-th/0208207].\\
G. Kofinas, E. Papantonopoulos and V. Zamarias, \textit { Black
Holes on the Brane with Induced Gravity}, Astrophys. Space
Sci. {\bf283} (2003) 685, [arXiv:hep-th/0210006].\\
G. Kofinas, \textit {General Brane Dynamics with $^{(4)}\!R$ term in
the Bulk}, JHEP {\bf0108} (2001) 034, [arXiv:hep-th/0108013].\\
G. Kofinas, E. Papantonopoulos and I. Pappa, \textit {Spherically
Symmetric Braneworld Solutions with $^{(4)}\!R$ term in the Bulk},
Phys. Rev. D {\bf66} (2002) 104014, [arXiv:hep-th/0112019].

\bibitem{grg}
R. Gregory, R. Whisker, K. Beckwith and C. Done, \textit {Observing
braneworld black holes}, JCAP {\bf0410}
(2004) 013, [arXiv:hep-th/0406252].\\
S. Creek, R. Gregory, P. Kanti and B. Mistry, \textit {Braneworld
stars and black holes}, Class. Quant. Grav. {\bf23} (2006) 6633,
[arXiv:hep-th/0606006].

\bibitem{gab}
G. Gabadadze and A. Iglesias, \textit {Schwarzschild Solution In
Brane Induced Gravity}, Phys. Rev. D {\bf72} (2005) 084024,
[arXiv:hep-th/0407049].

\bibitem{greg}
R. Gregory, \textit {Braneworld Black Holes}, Lect. Notes Phys. {\bf769} (2009) 259, [arXiv:0804.2595]. \\
R. Whisker, \textit {Braneworld Black Holes}, [arXiv:0810.1534]

\bibitem{dai}
D. -C. Dai and D. Stojkovic, \textit {Analytic solution for a static
black hole in RSII model}, [arXiv:1004.3291]

\bibitem{mid}
C. Middleton and G. Siopsis, \textit {The Schwarzschild solution in
the DGP model}, Mod. Phys. Lett. A {\bf19} (2004)
2259, [arXiv:hep-th/0311070].\\
E. Chang-Young and D. Lee, \textit {Charged Black Holes on DGP
Brane}, Phys. Lett. B {\bf659} (2008) 58, [arXiv:0708.3032].\\
D. Lee, E. Chang-Young, M. Yoon, \textit {Charged Rotating Black
Holes on DGP Brane}, Int. J. Mod. Phys. A {\bf24} (2009)
4389, [arXiv:0711.1074].\\
K. Zhang, P. Wu and H. Yu, \textit {The stability of Einstein static
universe in the DGP braneworld}, Phys. Lett. B {\bf690} (2010) 22,
[arXiv:1005.4201].

\bibitem{hey}
M. Heydari-Fard and H. R. Sepangi, \textit {Spherically symmetric
solutions and gravitational collapse in brane-worlds}, JCAP {\bf02}
(2009) 029, [arXiv:0903.0066].
 M. Heydari-Fard, H. Razmi and H. R. Sepangi, \textit {Brane-World
Black Hole Solutions via a Confining Potential}, Phys. Rev. D
{\bf76} (2007) 066002, [arXiv:0707.3558].\\
M. Heydari-Fard, \textit {Black hole solutions in warped DGP brane
world}, Astrophys. Space Sci. {\bf 325} (2010) 287.

\bibitem{vdvz}
Y. Iwasaki, \textit {Consistency Condition For Propagators}, Phys. Rev. D {\bf 2} (1970) 2255.\\
H. van Dam and M. J. G. Veltman, \textit {Massive And Massless
Yang-Mills And Gravitational Fields}, Nucl. Phys. B \textbf{22} (1970) 397.\\
V. I. Zakharov, JETP Lett. \textbf{12} (1970) 312.

\bibitem{vain}
A.I. Vainshtein, \textit {To The Problem Of Nonvanishing Gravitation
Mass}, Phys. Lett. \textbf{39B}, (1972) 393.

\bibitem{damour}
T. Damour, I. I. Kogan and A. Papazoglou, \textit {Spherically
symmetric spacetimes in massive gravity},  Phys. Rev. D \textbf{67},
(2003) 064009, [arXiv:hep-th/0212155].

\bibitem{mass}
A. Higuchi, Nucl. Phys. \textbf{B282}, (1987) 397; ibid. \textbf{B325}, (1989) 745.\\
I.I. Kogan, S. Mouslopoulos, and A. Papazoglou, \textit {The
$m\rightarrow0$ limit for massive graviton in dS$_4$ and AdS$_4$-How
to circumvent the van Dam-Veltman-Zakharov discontinuity}, Phys.
Lett. B \textbf{503}, (2001) 173, [arXiv:hep-th/0011138].

\bibitem{DGP2}
C. Deffayet, G. R. Dvali, G. Gabadadze and A. I. Vainshtein, \textit
{Nonperturbative Continuity in Graviton Mass versus Perturbative
Discontinuity},
 Phys. Rev. D \textbf{65}, (2002) 044026, [arXiv:hep-th/0106001].\\
A. Lue, \textit {Cosmic strings in a braneworld theory with metastable gravitons}, Phys. Rev. D \textbf{66}, (2002) 043509, [arXiv:hep-th/0111168].\\
M. Porrati, \textit {Fully covariant van Dam-Veltman-Zakharov
discontinuity, and absence thereof}, Phys. Lett. B \textbf{534},
(2002) 209-215, [arXiv:hep-th/0203014].

\bibitem{gruz}
A. Gruzinov, \textit {On the Graviton Mass}, New Astron. \textbf{10}
(2005) 311, [arXiv:astro-ph/0112246].

\bibitem{tanaka}
T. Tanaka, \textit {Weak gravity in DGP braneworld model}, Phys.
Rev. D \textbf{69}, (2004) 024001, [arXiv:gr-qc/0305031].

\bibitem{gariga}
J. Garriga and T. Tanaka, \textit {Gravity in the Randall-Sundrum
Brane World}, Phys. Rev. Lett. \textbf{84}, (2000) 2778,
[arXiv:hep-th/9911055].

\bibitem{gr}
Ruth Gregory, Raymond Laflamme, \textit {Evidence for the Stability
of Extremal Black p-Branes}, Phys. Rev. D \textbf{51}, (1995) 305,
[arXiv:hep-th/9410050].

\bibitem{dai2}
A. Satz, F. D. Mazzitelli and E. Alvarez, \textit {Vacuum
polarization around stars: Nonlocal approximation}, Phys. Rev. D
\textbf{71} (2005) 064001, [arXiv:gr-qc/0411046].\\
J. W. Moffat and V. T. Toth, \textit {Modified gravity and the
origin
of inertia}, MNRAS (2009) 395 , [arXiv:0710.3415].\\
D. -C. Dai, I. Maor and G. Starkman, \textit {Modified gravity:
living without Birkhoff I. DGP}, Phys. Rev. D \textbf{77} (2008)
064016 [arXiv:0709.4391].\\


\end{thebibliography}
\end{document}